# An Algorithm for Index Multimedia Data (Video) using the Movement Oriented Method for Real-time Online Services

Aries MUSLIM[(1)], A. Benny MUTIARA[(2)], Cut Maisyarah KARYATI[(3)], Purnawarman MUSA[(4)]

[1,3,4] **Etudient Doctoral Informatique Le2i - UMR CNRS 5158 Université de Bourgogne, 9 Av. Alain SAVARY, 21078 DIJON CEDEX**

[1,3]**Doctoral Student University of Gunadarma, Margonda Raya No. 100, Depok – Indonesia**

[2] **Faculty Computer Science Gunadarma University, Margonda Raya No. 100, Depok – Indonesia**

email : aries.muslim,purnawarman.musa[@etu.u-bourgogne.fr][1,4],
amuslim,amutiara,csyarah[@staff.gunadarma.ac.id][1,2,3]

## ABSTRACT

Multimedia data is a form of data that can represent all types of data (images, sound and text). The use of multimedia data for the online application requires a more comprehensive database in the use of storage media, Sorting / indexing, search and system / data searching. This is necessary in order to help providers and users to access multimedia data online. Systems that use of the index image as a reference requires storage media so that the rules and require special expertise to obtain the desired file. Changes in multimedia data into a series of stories / storyboard in the form of a text will help reduce the consumption of media storage, system index / sorting and search applications. Oriented Movement is one method that is being developed to change the form of multimedia data into a storyboard.

**Index Term-** Index Algorithm, Multimedia data, Movement Oriented.

## 1 INTRODUCTION

In the decades when the use of digital media is growing rapidly, both in size and type of data not only on the text but also on the image, audio and video. Along with the increasing use of digital media, especially video, is required technical data management and retrieval of image effectively. Volume of digital video produced by the field of scientific, education, medical, industrial, and other applications available for its users increased dramatically as a result of the progress in sensor technology, the Internet and new digital video. Engineering tools, annotated video with text and image search using the text-based approach. Through the description text, image can be organized by semantic hierarchy for easy navigation and search based on standard Boolean queries. Because, the description text for a broad spectrum of video can not be obtained automatically, the system mostly text-based video retrieval requires annotation manually. Indeed, manual video annotation is a work difficult and expensive for a large video database, and is often a subjective, context-sensitive and is not perfect. (Long Fuhui, Chia-Hung, Hove, 2005).

There are two approaches that can be used to represent the video: (1) metadata-based and (2) Content-based. Techniques required for the retrieval (query) from the two approaches which can be divided into 3 namely: (1) Context-based, (2) Semantic-based and (3) Content-based.

Content-based video retrieval, a technique that uses cargo for visual image of a large-scale video database according to user needs, has developed rapidly since the 1990s. Content based image retrieval, using the cargo from a visual image or video such as color, shape, texture, and spatial layout to provide indexing and image or video. In the system content-based image retrieval is typically shown in Figure 1, to recover image / video, the user provides an example of the image / video retrieval as a reference. System and then change the image in this example to continuous feature vector. Similarities between the feature vector of an example query image in the database and then calculated and retrieval is done with the help of an index scheme. Index scheme provides a way to efficiently find the image database. Retrieval system has the latest feedback of the user to modify the retrieval process in order to produce retrieval results that are fuller of meaning and perception semantics.

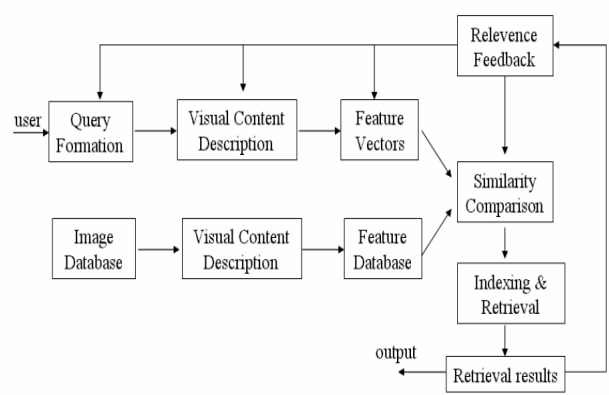

Figure 1 System Diagram Content-Based Image / Video Retrieval

## 2 OVERVIEW AND REFERENCES

Multimedia (video) have 3 data elements, namely, 1).The Image, 2). The voice 3). Text or character of a display to other combinations of these three elements ( Nalin Sharda, 2005). Multimedia system design is a very complex problem.

Where is the system complexity is growing a lot of elements on each linier changes or nonlinear of the results presentation and interaction system. Design oriented movement / Movement Oriented Design is a new paradigm that will help manage the level of complexity in the design problem and search multimedia data (video) is so diverse (Nalin Sharda, 2005).

Data / video retrieval on the number of video database with a very large system using content-based experience will problem Identify the time to give each frame of video files, which is a series of Identify textual each data object that will represent the video frame, where a combination of data is a metadata that will add more media storage needs that are not less (William I. Grosky, 1997). Textual process of adding data to the multimedia data such as figure 2

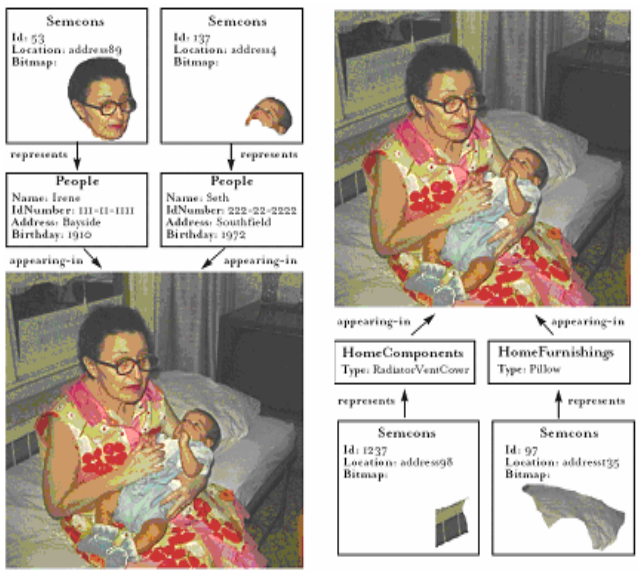

Figure 2. Content Base Method (Semcon) Identify Multimedia Data

Design-oriented movement / Movement Oriented Design is a basic concept of the form of multimedia system, whether they are formal (educational) or a game (game), all of which have a certain flow of the story itself. Art establishment flow must be a story of integration of 3 scientific disciplines, namely 1) art and story telling ability / art story telling, 2) Cognitive psychology and 3) technology / programming. (Nalin Sharda, 2005) At the time of this object-oriented program is a key tool in making the system a multimedia / video that you can manipulate the object of a story, either in part or whole flow of multimedia stories. Changes can be made based on the interpretation of the respective manufacturer / user of a series of multimedia stories. These changes are very dependent from the level of knowledge, experience, and psychological situation of the personnel to do so. This design will produce a series of stories that come from each multimedia frame that will form the collection of data / database textual which is the identification of each frame of the overall display multimedia data. Collection of data is the data textual will alphanumeric that easy to do with the process and indexing query during the search process the data.

## 2.1 Problem

1. Collection story will multimedia representation of the set of statements / character in the form of a database is relatively large because of the various capabilities of each of the personal view (Nalin Sharda, 2005). This is like the picture 3.
2. Type the story of multimedia data will reflect the distinctive character, which of course will vary according to table 1.
3. Indexing form of navigation and data processing work will require a relative faster, because each one (1) frame / stage will result in some textual statement, which requires the index of the optimal algorithm of data collection of stories, when done the search process or queries. As an example in table 2 Figure 1. Integration of directory server

Table 1. The characters in the table need a multimedia data

| Story Type | Charakter Type |
|---|---|
| Humanistic | Human Beings |
| Animated | Animation Beings |
| Game | Game Beings |
| Education | Knowledge elements |
| Song | Word, metaphors |
| Musik | Notes, Movements |
| Multisensory story | " + Touch, Smell & Taste |
| Formal Story | Any of the above |

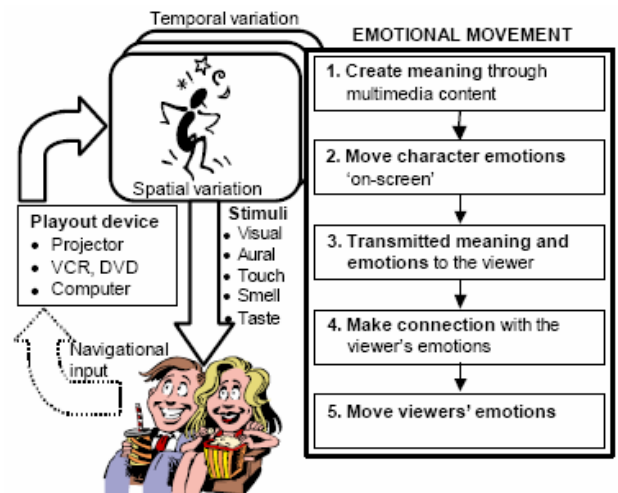

Figure 3. Multimedia Data transformation in the form of Textual Data

**Tabel 2 Main Problem: To explain and demonstrate important aspects of electric current to high school students.**

**STAGE-1**
**Problem: To explain the meaning of electric current**
**B1** Importance of Electric Current (EC)
**M1** Define and exemplify EC
**E1** Link to Ohm's Law. Explain AC & DC Each of the Stage-1 Story Units, i.e.
**B1, M1** and **E1** can now be expanded further. For example, '**B1,B2**' in Stage-2 is the Begin of the Story Unit '**B1**' in stage-1.

**STAGE –2**
**B1 Problem: Why is EC important**
*Explain the importance of Electric Current (EC)*
**B1,B2** Many people die of electric shock
**B1,M2** Understand and respect EC, not be afraid of it.
**B1,E2** EC is useful for running appliances
**M1 Problem: How is EC defined**
*Define and exemplify EC*
**M1,B2** Amperes = Coulombs / second
**M1,M2** It's like watching Coulombs go past and counting how many go past in one second.
**M1,E2** Demonstrate the effect of EC through multimedia and *multisensory experience*.
**E1 Problem: What determines EC strength**
*Link to Ohm's Law. AC/DC*
**E1,B2** Current depends upon voltage and resistance
Coulombs go past and
counting how many go past in one second.
**M1,E2** Demonstrate the effect of EC through multimedia and *multisensory experience*.
**E1 Problem: What determines EC strength**
*Link to Ohm's Law. AC/DC*
**E1,B2** Current depends upon voltage and resistance
E1,M2 Ohm's Law: I = V/R
**E1,E2** Current can be direct or alternating.
Some of the Story Units at Stage-2 have been expanded further in Stage-3; whereas some of these have been left out, signifying that Story Units can be instantiated in more ways than one.

**STAGE -3**
**B1 Problem: Why is EC important**
*The importance of Electric Current (EC)*
**B1, B2 Problem: Effect and cause of electric shock**
*Many people die of electric shock*
**B1,B2,B3** Video clip of a person getting a shock
B1,B2,M3 Explain the reason for the shock
B1,B2,E3 Ask, "So what is electric current?"
**B1, M2 Problem: How should we treat EC**
*Understand and respect EC, not be afraid of it.*
**B1,M2,B3** (Left un-instantiated
B1,M2,M3 for the user to
B1,M2,E3 try out some options)
**B1, E2 Problem: How do we use EC**
*EC is useful for running appliances*
B1,E2,B3 (Left un-instantiated
B1,E2,M3 for the user to
B1,E2,E3 try out some options)
**M1 Problem: How is EC defined**
*Define, exemplify and inject EC*

## 3 RESEARCH METHOD

Refer to the problems in the process of change to the multimedia data in the textual data with the movement oriented method, which results in a textual database as multimedia Identify data that will be in the index. So the steps that the research will be conducted is:

1. Gathering Video Data-based educational teaching about Robot (MPG, AVI , MPEG3 / 4)
2. Editing / Cutting of video data objects based on the main story (Video Editing, Adobe Premiere / Ulead Video Studio 7 / 8)
3. Build a database metadata (story board, image and sound) which comes from the editing / cutting data to determine tuple video that will be the object index
4. Build a video database with object oriented programming architecture based on ACOI (A System for Indexing Multimedia Object) to get tuple to process the index.
5. Determining the stage for a major input in the feature detector Detection Engine.
6. Make a simulation with the available databases, to design algorithms index of the overall data.
7. To measure the success of the design algorithm index, conducted the simulation / testing query data on the original database with a database that has been in the index, when DBA T <T dBi, then the index is not successful, the algorithm index enhanced DBA Q>> Q dBi, then the index is successful, the process method with the index movement oriented to the multimedia data can be done.

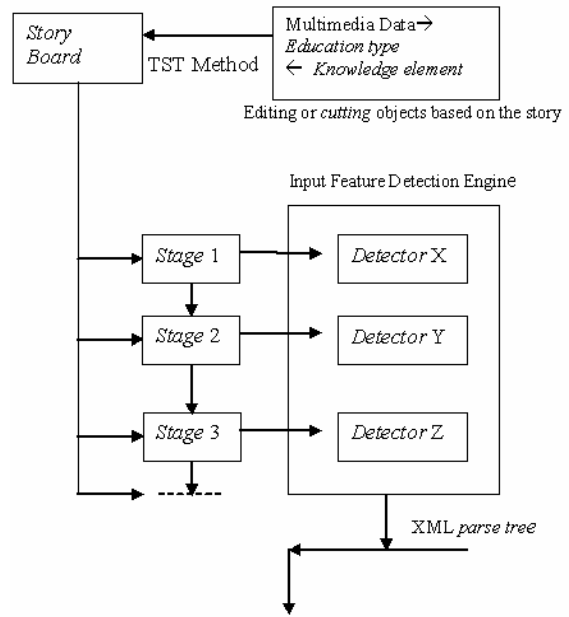

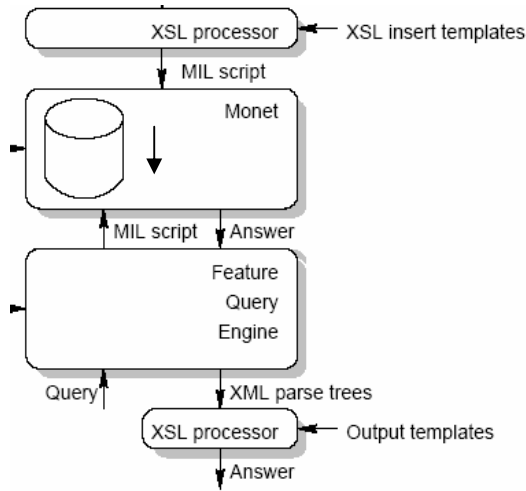

Figure 4 Diagram Process data fragmentation Video

Movement Oriented A new paradigm for multimedia design, called the Movement Oriented Design will change frame-per-frame data into a series of multimedia story (text only / stage) based on the movement, making the process Identify and index and search data to be more simple, this is because the data will be textual data exploration is the ideal form of notice is required by a database system, whether small or large.

Multimedia data that education has a specification of a particular approach so that changes every frame of the story plot does not become too diverse. This will facilitate the process to Identify image and voice into a series of story / describing the stage of multimedia data.

Identify a frame that has been based textual data will be easier to classified based on the limitations of the multimedia data. For example learning about the movement of robots, the classification can be drawn is that, 1) type of robot, 2) the form of a robot, 3) type of movement, 4) a lot of movement, without having to see the background, about the condition or feelings creator element. Object Oriented Programming is a technique that supports the process of classification into objects that can be used as a restriction of the index data.

Quality Of Service (QOS) is the important part that must be a reference Identify when the multimedia data into textual data does not have the same level of knowledge of each personal, so that interpretation is not a frame out of the path of operation. (Melanie Anne Phillip, Chirs Huntley, 2001)

Algorithm for Parallel Process Video The / data real-time video with search range 40 ms / period requires a technique that is very complex, where the frame of an object based on the main level partition needs. (Altilar DT, Paker Y, 2001). Methods partition the video data that have been developed, namely, (1) AST - Almost Square Tiles data partitioning algorithm, (2) AST war - Almost Square Tiles data partitioning algorithm with aspect ratio, and (3) Lee-Hamdi.

Following algorithm is AST

1) k ← least greater or equal square(partitions)
2) first square ← squareroot(k) (A)
3) cols ← first square
4) if (partitions is a square of an integer) rows ← first square (B)
else if ((rows-1)*cols partitions) rows ← first square -1
5) irr col ← cols * rows - partitions (C)
6) a ← image height/rows
7) ap ← image height - a * (rows -1)
8) ar ← image height/num rows - 1 (D)
9) arp ← image height - ar * (rows - 2)
10) b ← (image width/((ar/a) * (cols-irr cols) + irr cols)) * (ar/a)
11) bp ← (image width - b * (cols-irr cols) ) / irr cols (E)
12) bpp ← image width - b * (cols-irr cols) - bp * (irr cols - 1 )

RST - regular standard ones ($a \quad b$),
RET - regular excess ones ($ap \quad b$),
IST - irregular standard one ($bp \quad ar$),
ICET - irregular column excess ones ($bpp \quad ar$),
IRET - irregular row excess ones ($bp \quad arp$),
IRCET - irregular double excess one ($bpp \quad arp$);

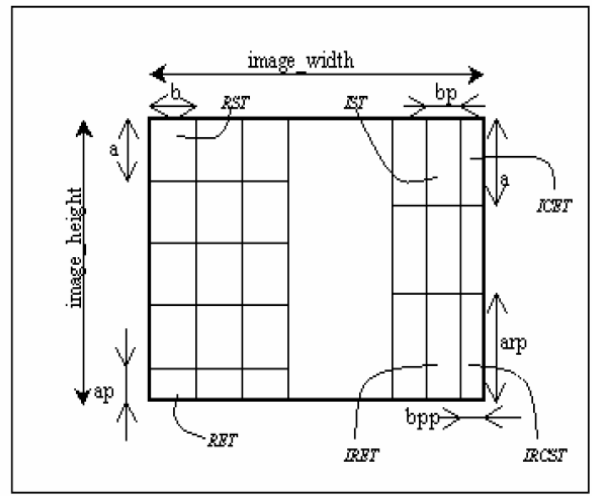

Figure 5. Picture of the position of video data on Algorithm AST

From the discussion above that the algorithm developed will be very difficult if the video database in large numbers, because will be very many objects to which the process of identification constrain index.

Indexing Multimedia Objects

Creating Index-based multimedia objects, which provides freedom in the grammar will form a new structure on the semi-main memory database system, using the derived parse tree to make the index on the source of multimedia data (Menzo Windhouwer, Albrecht Schmidt, Martin Kersten, 2000)

Detector X represent the voice data, the detector is variable among others;
1) Voters people based on gender, region of origin, age, and so forth. 2) Hearing music: music tabuh instruments, stringed instrument, musical instrument quotation, 3) natural Voters, andothers.

Detector Y represent the image data, with variable detector, among others:
1) Form of the picture, 2) type of image, which can be detected with a color histogram, edge detectors and so forth. Air Z is the representation of text data, the variables detector more similar.

Of the above conditions will result in the diversity of feature detection, but can be overcome by using the parse tree-based object, such as the example below:

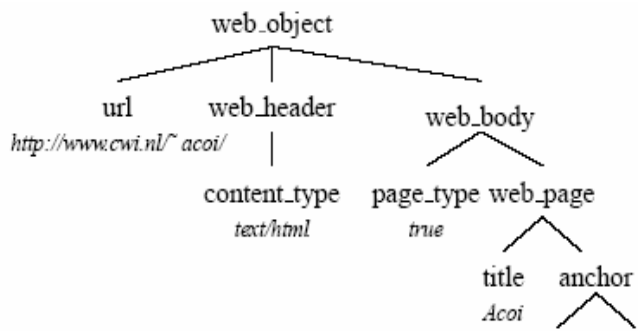

Figure 6. Parse Tree Based Object

From the parse tree approach will produce a tuple is unique, so it can be used as a variable index of the multimedia database.

## 4   CONCLUSION AND DISCUSSION

### 4.1 Conclusion

The division of a series of frames from video data movement, can not be done in a partial, to avoid errors in interpretation of the relationship frame with other frames
The process of change in the video frame into a set of stories that shaped the text, a very personal knowledge each of which translated, so that the necessary standardization knowledge on the system that was built.
The process of cutting the frame to become the most crucial to the next step, because the cutting process is expected to be automated, (currently still rely on manual with the personal ability)
Identification of sound and image in the process of development methods, which may be able to collaborate with other research

### 4.2 Discussion

Development of QoS, a priority for further development, this standard provides for the cutting **measurement** the right
To measure the success of the design algorithm index, conducted the simulation / testing query data on the original database with a database that